\documentclass[twocolumn]{aastex6}
\usepackage{apjfonts}

\begin{document}

\title{Assembling the Milky Way bulge from globular clusters: Evidence from the double red clump}

\author{
Young-Wook Lee\altaffilmark{1,}\footnotemark[3],
Seungsoo Hong\altaffilmark{1,}\footnotemark[3],
Dongwook Lim\altaffilmark{1,}\footnotemark[3],
Chul Chung\altaffilmark{1},
Sohee Jang\altaffilmark{1},
Jenny J. Kim\altaffilmark{1}, 
Seok-Joo Joo\altaffilmark{2}
}

\altaffiltext{1}{Center for Galaxy Evolution Research \& Department Astronomy, Yonsei University, Seoul 03722, Korea; ywlee2@yonsei.ac.kr} 
\altaffiltext{2}{Korea Astronomy and Space Science Institute, Daejeon 34055, Korea}


\begin{abstract}
The two red clumps (RCs) observed in the color-magnitude diagram of the Milky Way bulge is widely accepted as evidence for an X-shaped structure originated from the bar instability.
A drastically different interpretation has been suggested, however, based on the He-enhanced multiple stellar population phenomenon as is observed in globular clusters (GCs). 
Because these two scenarios imply very different pictures on the formation of the bulge and elliptical galaxies, understanding the origin of the double RC is of crucial importance.
Here we report our discovery that the stars in the two RCs show a significant ($>$ 5.3$\sigma$) difference in CN-band strength, in stark contrast to that expected in the X-shaped bulge scenario. 
The difference in CN abundance and the population ratio between the two RCs are comparable to those observed in GCs between the first- and later generation stars. 
Since CN-strong stars trace a population with enhanced N, Na, and He abundances originated in GCs, this is direct evidence that the double RC is due to the multiple population phenomenon, and that a significant population of stars in the Milky Way bulge were assembled from disrupted proto-GCs. 
Our result also calls for the major revision of the 3D structure of the Milky Way bulge given that the current view is based on the previous interpretation of the double RC phenomenon.
\end{abstract}

\section{Introduction}\label{intro}
\begin{figure*}
\centering
\includegraphics[width=0.90\textwidth]{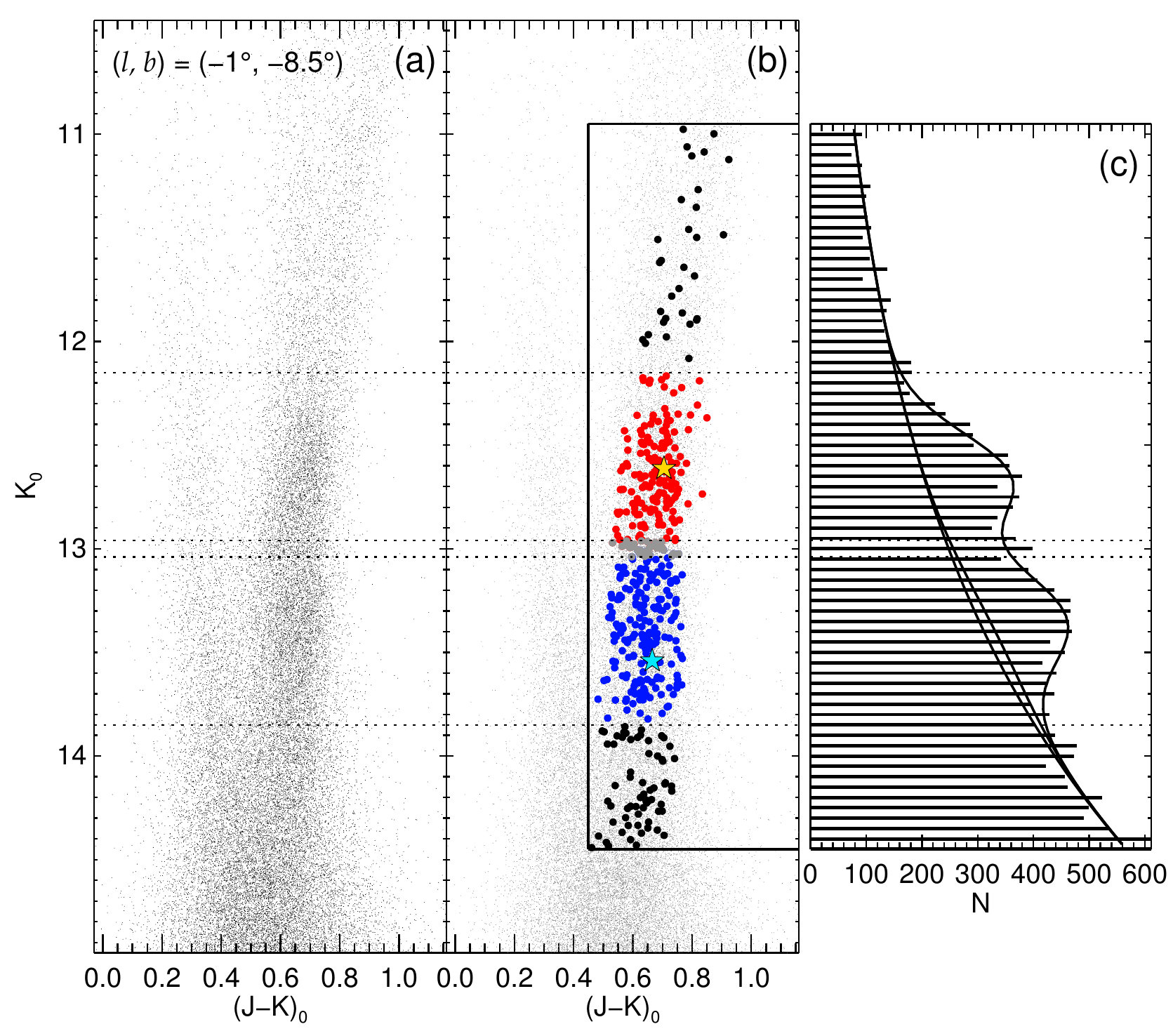}
\figcaption{
Near-IR CMD of our bulge field. (a) All stars in our bulge field from 2MASS survey. (b) The red, blue, and black circles indicate our sample stars in bRC, fRC, and RGB regimes selected for spectroscopic observations. The star symbols are for two example spectra shown in Figure~\ref{fig_sed}. The CMD is contaminated by the foreground disc main-sequence stars at (J-K)$_{0}$ $<$ 0.5, and therefore the selection box for the bulge stars is displayed by the black lines. (c) The double RC is clearly visible in the luminosity function shown with fitting functions \citep{ref29}. The population ratio between bRC and fRC is 0.52 ($\pm$0.01) : 0.48 ($\pm0.01$).
\label{fig_cmd}
}
\end{figure*}

\footnotetext[3]{These authors contributed equally to this paper.}
The red clump (RC) stars are the metal-rich counterpart of core-He-burning horizontal-branch (HB) stars, and are ubiquitously observed in color-magnitude diagram (CMD) of the Milky Way bulge. 
In the high-latitude ($|$b$|$ $>$ 6$^{\circ}$) field of the bulge, the RC is surprisingly split into two groups, with a magnitude difference of $\sim$0.5 between the bright and the faint RCs (bRC, fRC; see Figure~\ref{fig_cmd}). 
Based on the latitude and longitude dependence of this double RC feature and the presence of an X-shaped structure in some extragalactic bulges, it was originally interpreted as evidence for a giant X-shaped structure in the Milky Way bulge originated from the buckling instability of a bar \citep{ref1,ref2,ref13,ref12}. 
According to this scenario, the foreground (background) arm of the X-structure would be observed as bRC (fRC) to the observer on the earth. 
While it is well established that the low-latitude field of the bulge is dominated by a bar population, this led to the consensus that even the high-latitude field of the bulge has ``pseudo bulge'' characteristic. 
However, a completely different interpretation has been suggested \citep{ref3,ref4,ref5} motivated by the multiple stellar population phenomenon observed in GCs \citep[][and references therein]{ref6,ref7,ref8,ref9}. 
In this model, the bRC comprises He-enhanced second- and later generation (G2) stars, while the fRC is composed of He-normal first-generation (G1) stars\footnote{For this interpretation, they used synthetic HB models assuming two populations with very different He abundances. The required difference in He content in the metal-rich bulge is much larger than those generally observed in metal-poor GCs between G2 and G1. However, this apparently strong metallicity dependence of He enhancement between G2 and G1 is predicted in a chemical evolution model \citep{ref22} because  the He content within the winds of massive stars increases with metallicity \citep{ref23,ref103}. Some support for this is provided by synthetic HB models for metal-rich bulge GCs, including NGC 6388 and Terzan 5 \citep{ref101,ref201,ref5}.}.
In GCs, He-enhanced G2 stars are also enhanced in N and Na \citep{ref7,ref11}.
\par
In GCs with multiple populations, the CN-band traces N abundance, and therefore it has been widely used in the search for stars originated from GCs in the halo and the inner Milky Way \citep{ref15,ref16}. 
This chemical tagging based on CN-band strength is now well established and can also be used in the investigation of the origin of the double RC phenomenon. 
The purpose of this Letter is to report our discovery that the stars in the two RC regimes show significant difference in the CN-band strength, supporting the multiple population origin of the double RC phenomenon. 
\par

\section{Observations and data analysis}\label{obs}
We have observed CN-band strengths for 478 RC and red giant branch (RGB) stars in the high-latitude bulge field, (l, b) $\sim$ (-1$^{\circ}$, -8.5$^{\circ}$), where the double RC is most prominent. 
Figure~\ref{fig_cmd} shows our sample stars in RC and RGB regimes in the J-K vs. K CMD obtained from 2MASS All-Sky Point Source Catalog \citep{ref30}. 
Note that our RC zones in CMD are mixed with background RGB stars, where the RC and RGB stars are spectroscopically indistinguishable \citep{ref18,ref17}. 
We will therefore compare the mean difference in CN-band strength between the two RC zones, which can be then converted to the difference between only the genuine RC stars. In order to minimize the effect from evolutionary mixing, we have restricted the RGB stars to within $\pm$1.2 V mag ($\pm$2.0 K mag) from the RC level.
\par
\begin{figure}
\centering
\includegraphics[width=0.5\textwidth]{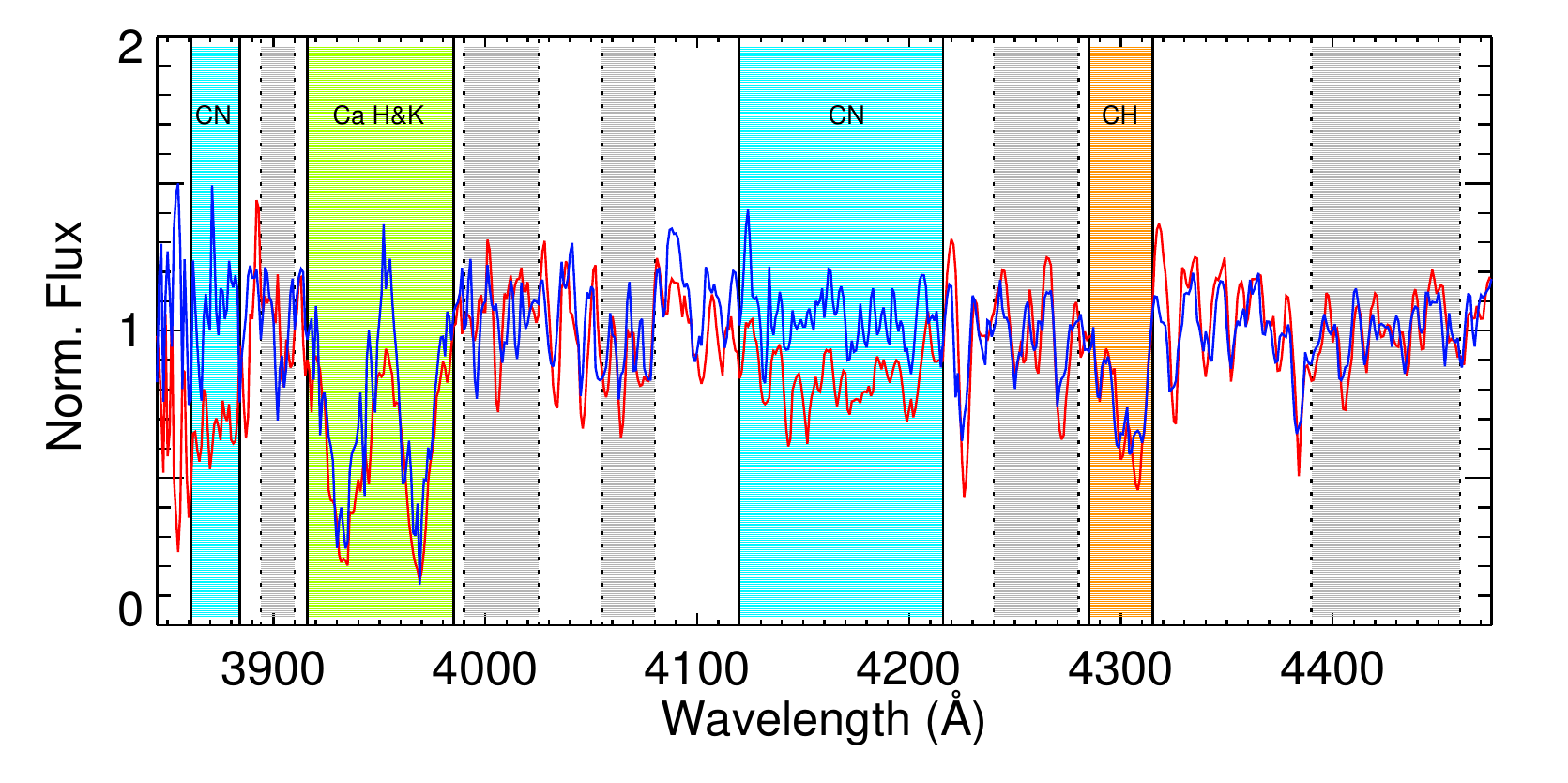}
\figcaption{
Example of the observed spectra for the two stars in bRC and fRC. Two spectra, each representing stars in the bRC (red) and fRC (blue) regimes (see the star symbols in Figure~\ref{fig_cmd}), are compared. The CN, CH, and Ca H\&K bands are indicated with their continuum bands in grey. The S/N for these spectra, measured at $\sim$4000 {\AA}, is $\sim$ 23.
\label{fig_sed}
}
\end{figure}

Our low-resolution spectroscopy was carried out with the same instrument setup and data reduction procedures used for our previous investigations of multiple populations in GCs \citep{ref10,ref19}. 
The observations were performed with the du Pont 2.5-m telescope at Las Campanas Observatory in 2016 June and 2017 April and June. 
We used WFCCD instrument with HK grism providing a 25{\arcmin} $\times$ 25{\arcmin} field-of-view with a dispersion of 0.8 {\AA}/pixel and a spectral coverage 3,500 -- 5,000 {\AA}. 
For the program field, 16 multi-slit masks were generated, each of which includes about 35 slits of 1{\arcsec}.2 width. 
Typically, three 1800 s science exposures, three flats, and an arc lamp frames were taken for each mask. 
The raw data were reduced using IRAF\footnote{IRAF is distributed by the National Optical Astronomy Observatory, which is operated by the Association of Universities for Research in Astronomy (AURA) under a cooperative agreement with the National Science Foundation.} and the modified version of WFCCD reduction package in the usual manner \citep{ref31,ref10}.
In addition, we performed radial velocity correction with the IRAF RV package.
Figure~\ref{fig_sed} shows two typical examples of spectra from our observations, each representing stars in the bRC and fRC regimes, respectively. 
\par
The strengths of CN, CH molecular bands and Ca H\&K lines were estimated from the commonly used spectral indices, CN(3839), CN(4142), CH(4300) and HK$'$ \citep{ref14,ref32,ref10}.
Following the same procedures we adopted for RGB stars in GCs in our previous studies, we measured these indices individually for each spectrum as the ratio of absorption strength to the nearby continuum. 
The measurement errors were calculated assuming Poisson statistics in the flux measurements. 
The final indices for a star were obtained by taking the error-weighted mean of the indices measured from each exposure excluding low quality spectra with signal-to-noise ratio (S/N) at 3900 {\AA} $<$ 5. 
In addition, stars with large measurement error or those abnormally far from the mean of the distribution are rejected to a confidence level of 99.7$\%$. 
A total of 478 stars (176 for bRC; 206 for fRC; 96 for RGB) were selected for our analysis, which are identified in Figure~\ref{fig_cmd}. 
We also measured delta indices, $\delta$CN(3839), $\delta$CN(4142), $\delta$CH(4300), and $\delta$HK$'$, as the difference between the original index and the least-square lines for RGB stars, to see the difference of chemical composition without the effects of effective temperature and surface gravity.
 Finally, we calculated the mean differences of $\delta$-indices between the stars in the bRC and fRC groups. 
 The confidence level of the difference was estimated from the error of the mean for each group.
\par

\begin{figure}
\centering
\includegraphics[width=0.5\textwidth]{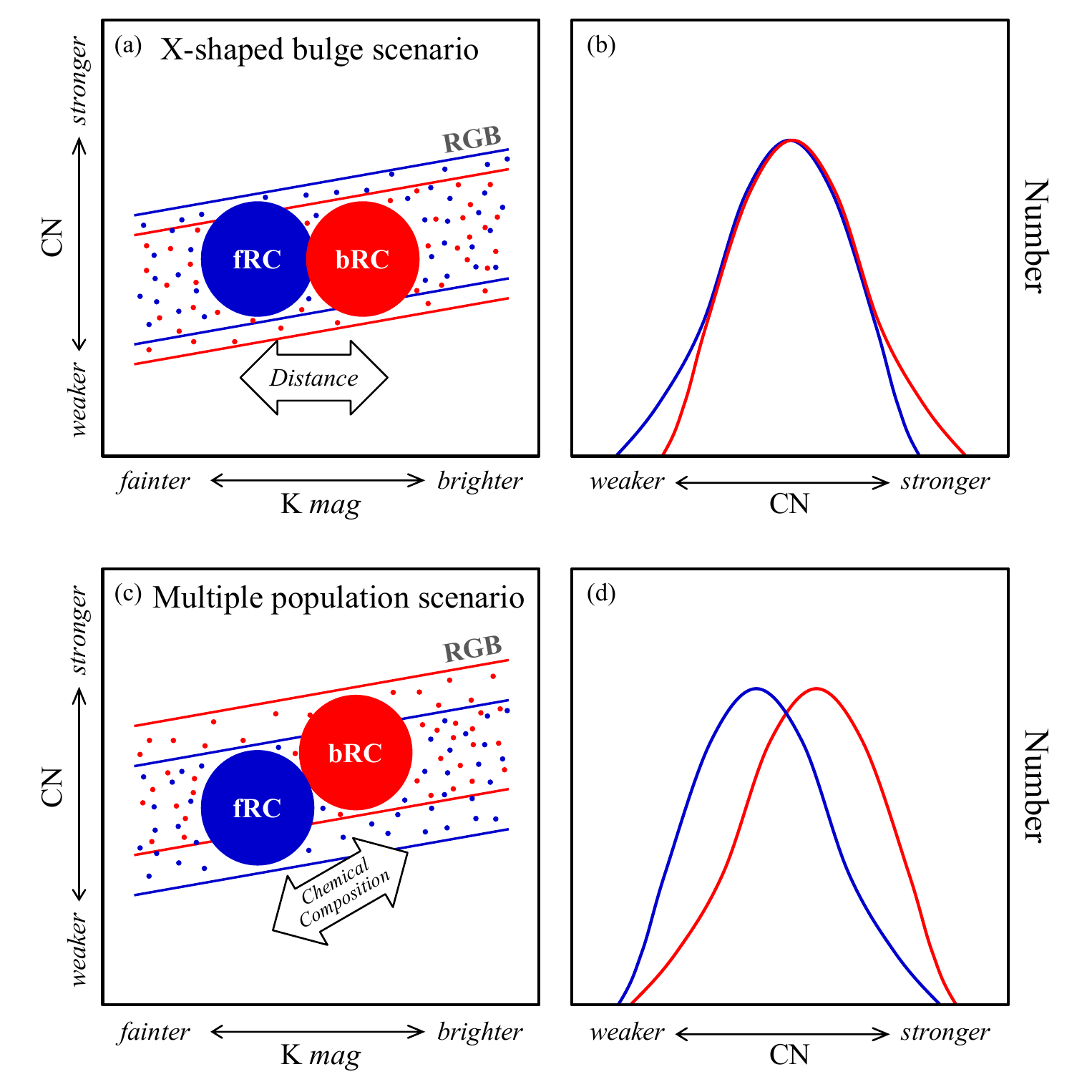}
\figcaption{
Schematic diagrams for the predicted placements of fRC and bRC stars in K magnitude vs. CN index diagram. (a-b) In an X-shaped bulge scenario, we expect only a horizontal displacement of the bRC (red) compared to the fRC (blue), with no difference in the peak position of the CN index distribution between the stars in the bRC and fRC regimes. The dotted bands are for the background RGB stars of the corresponding subpopulations. (c-d) In the multiple population model, the stars in the bRC are brighter because of enhanced He, and are CN-strong because of N enhancement. Therefore, the stars in the bRC and fRC regimes are predicted to show a marked difference in CN index distribution.
\label{fig_schematic}
}
\end{figure}

\section{Multiple population origin of the double Red Clump}\label{result}
\begin{figure*}
\centering
\includegraphics[width=0.90\textwidth]{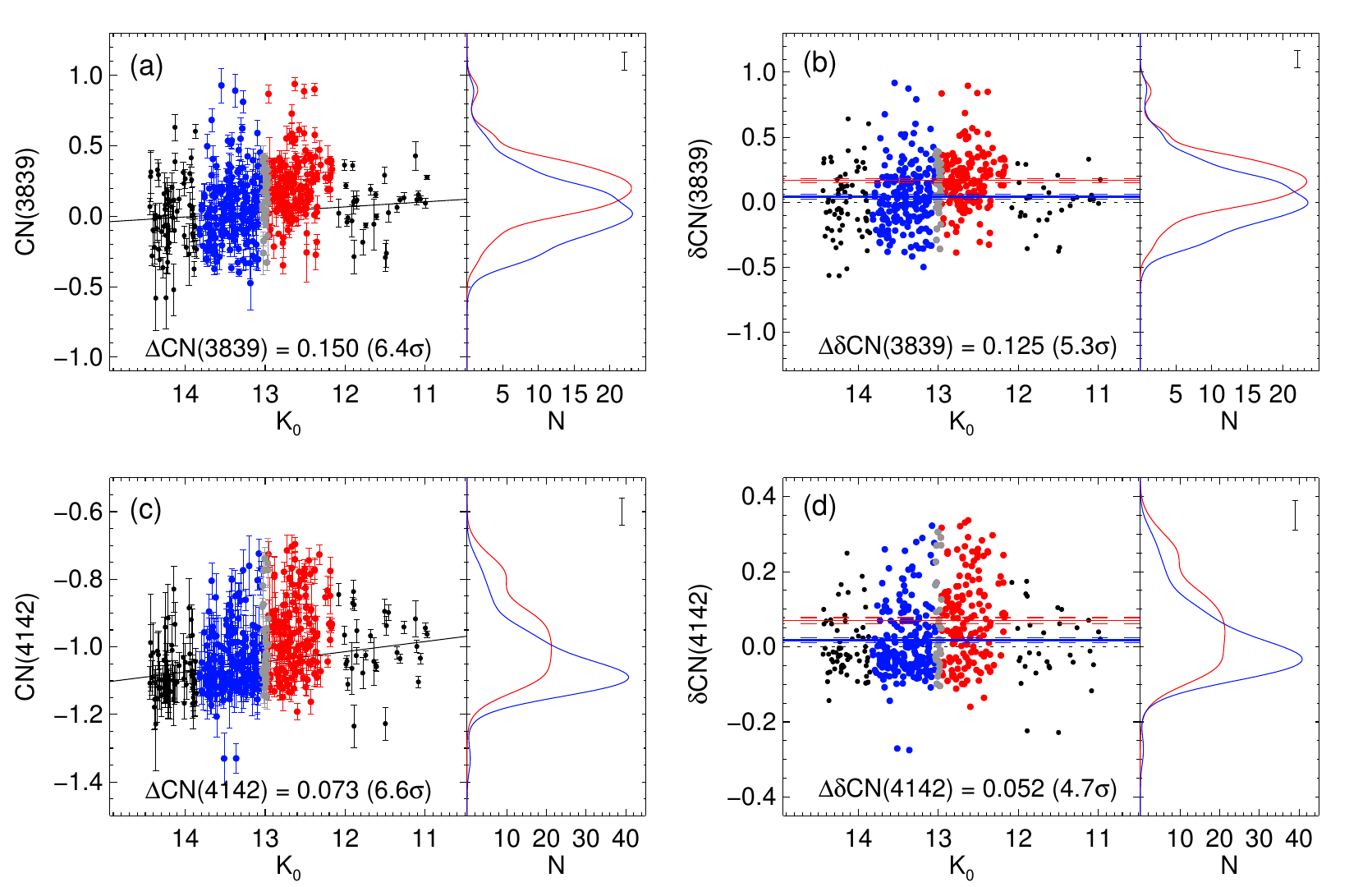}
\figcaption{
Measured CN indices for our sample stars in the bulge field. (a) CN(3839) index as a function of K magnitude. The red, blue, and black circles are stars in the bRC, fRC, and RGB regimes, respectively. The right panel compares CN index distributions of stars in the bRC (red) and fRC (blue) regimes, where the vertical bar indicates a Gaussian kernel width used in the generalized histogram. (b) The $\delta$-index is obtained from the height of CN index above the least-square line for RGB stars in both sides of the RC zone (black solid line in panel (a)). The solid and dashed lines denote, respectively, the mean value and its 1$\sigma$ error for each subpopulation. Note that the stars in the bRC regime are CN-enhanced compared to those in the fRC regime. (c-d) Same as panels (a-b), but for the CN(4142) index. Note again that the stars in the bRC regime are relatively CN-strong compared to those in the fRC regime.
The data used to create this figure are available.
\label{fig_cn}
}
\end{figure*}

\begin{figure*}
\centering
\includegraphics[width=0.90\textwidth]{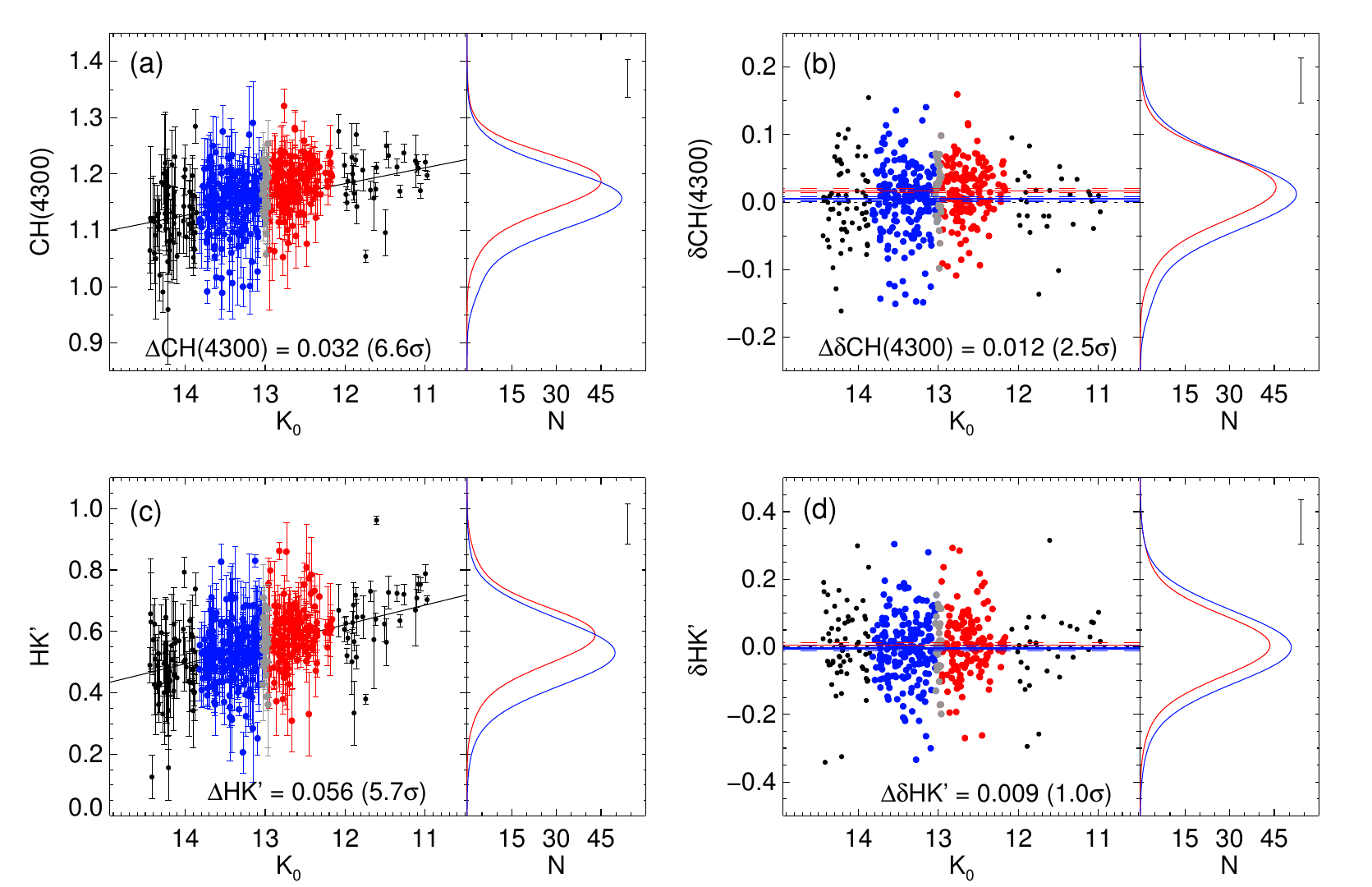}
\figcaption{Same as Figure~\ref{fig_cn}, but for CH and Ca HK$'$ indices. (a-b) Unlike CN-band, only very little difference is observed for $\delta$CH between the stars in the bRC and fRC regimes. (c-d) The difference in Ca abundance is negligible between the stars in two RC zones, consistent with the findings of previous investigations \citep{ref46,ref47}.
\label{fig_ch}
}
\end{figure*}
As illustrated in Figure~\ref{fig_schematic}, the two scenarios on the origin of the double RC predict completely different distributions of fRC and bRC stars in the magnitude vs. CN index diagram. 
In the case of an X-shaped bulge scenario, the difference in magnitude between the two RCs is due to the difference in distance, and the stars in the bRC would be identical with those in the fRC in terms of chemical composition. 
We therefore expect only a horizontal displacement of bRC compared to fRC in this diagram, with no difference in the peak position of the CN index distribution between the stars in the bRC and fRC regimes. 
On the other hand, in the multiple population model, the stars in the bRC would be G2 enhanced in N, He, and Na, while those in the fRC are G1 without these enhancements.
In this scenario, the stars in the bRC are brighter because of enhanced He \citep{ref3,ref5}, and are CN-strong because of N enhancement. 
Therefore, the stars in the bRC and fRC regimes are predicted to show a marked difference in CN index distribution. 
Note further that the direction of skewness of the distribution, for both bRC and fRC regimes, respectively, is just the opposite of that predicted in the X-shaped bulge scenario.
\par
The two CN indices measured, CN(3839) and CN(4142), are presented in Figure~\ref{fig_cn} in the K magnitude vs. CN index diagram. 
It is most obvious from the upper panels that stars in the bRC regime are systematically CN-enhanced compared to those in the fRC regime, just as expected in the multiple population scenario. 
The mean difference in the CN(3839) index between the stars in the bRC and fRC regimes is measured to be $\Delta$CN(3839) = 0.150 $\pm$ 0.023 mag, which is significant at the 6.4$\sigma$ level. 
For the delta index ($\delta$CN), we find more than 5$\sigma$ difference in $\delta$CN(3839) between the stars in bRC and fRC regimes ($\Delta\delta$CN = 0.125 $\pm$ 0.023 mag). 
A KS test also confirms that the probability is extremely low (0.00001$\%$) for the two samples to be drawn from the same distribution. 
When the relative fractions of background RGB stars (66\% and 78\% for bRC and fRC, respectively), which would belong to both G1 and G2, are taken into account (see Figure~\ref{fig_cmd}), this observed quantity would correspond to $\Delta\delta$CN $\approx$ 0.458 mag for the mean difference between only genuine RC stars in the two RC zones. 
We note that this difference is comparable to those observed in GCs between G1 and G2 \citep{ref10,ref19}. 
In addition, the CN index distribution is observed to be skewed to higher CN with a peak at lower CN for fRC, while it is skewed to lower CN with a peak at higher CN for the bRC. 
This is fully consistent with the multiple population scenario, but is just opposite of the predicted distribution of the X-shaped bulge scenario.
\par
A similar result is obtained from CN(4142), further strengthening the difference in $\delta$CN between the two RC samples. 
Unlike the behavior of the CN-band, only very little difference is observed for the CH-band ($\Delta\delta$CH = 0.012 $\pm$ 0.005 mag; see upper panels of Figure~\ref{fig_ch}).
This is qualitatively consistent with the trend observed in GCs, although, in the metal-rich regime, CN-strong stars are not observed to be extremely CH-weak as in metal-poor GCs \citep{ref15,ref18}. 
Furthermore, our observations show that the difference in heavy element abundance, as traced by Ca, is negligible ($\Delta\delta$HK$'$ = 0.009 $\pm$ 0.010 mag; see bottom panels of Figure~\ref{fig_ch}) between the stars in the two RC zones. 
This chemical pattern (CN-strong, CH-weak, and Ca-weak) is exactly the behavior we would expect from G2 stars with GC origin\footnote{In order to further confirm the GC origin of these stars, high-resolution spectroscopy would be required for Na and O abundances. The luminosity difference between the two RCs, however, could be considered as a sign of different He content, another genuine indicator of G2.}. 
One may argue that some difference in CN-band strength might be observed in the X-shaped bulge scenario as well, because the line of sight would cross different distances above the Galactic plane for the foreground and background arms of the X-structure. 
However, even if the X-shaped structure follows a shallow metallicity gradient observed in the bulge \citep[$\sim$0.04 dex/$^{\circ}$;][]{ref20}, the predicted difference in CN index from this effect alone is estimated to be almost negligible ($\sim$~5$\%$) compared to the observed difference between the two RC regimes.
\par

\section{Discussion}\label{discussion}
It is now well established that a significant number of CN-strong (and relatively CH-weak) stars can form only in a GC environment, and these N-rich G2 stars are also enhanced in He and Na \citep[][and references therein]{ref15,ref7}. 
Therefore, our result is direct evidence that the double RC is due to the multiple population phenomenon, and has little to do with an X-shaped structure. Stellar evolution models naturally predict super-He-rich stars to be placed on the bRC in the metal-rich population like the bulge \citep{ref3,ref4,ref5}, which is also seen in the metal-rich bulge GC Terzan 5 \citep{ref21,ref5}. 
Chemical evolution models also suggest a large difference in He abundance between G2 and G1 in the metal-rich regime \citep{ref22}, which is mostly due to the strong metallicity dependence of He yields from the winds of massive stars \citep{ref23}.
\par
The astonishing point from our result is an apparent similarity in the population ratio of G1 and G2 between our high-latitude bulge field and GCs\footnote{Again, we assume a strong metallicity dependence of He enhancement between G1 and G2 as predicted by a chemical evolution model of \citet{ref22}, where the ``mass budget problem" \citep{ref104} is much alleviated without the significant preferential loss of G1 suggested by \citet{ref102}.}.
This would indicate that a significant population of stars in the Milky Way bulge were assembled from disrupted proto-GCs or subsystems similar to GCs in terms of chemical evolution. 
Despite many uncertainties, it appears that GCs played a far more important role in the formation of the bulge than the outer halo, where only $\sim$3$\%$ of the stars show G2 characteristics \citep{ref15}. 
Chemical tagging with APOGEE survey has also discovered a large population of N-rich stars in the inner Galaxy \citep{ref16}. 
While this result is qualitatively consistent with our discovery of CN-enhanced stars in the bulge, a quantitative comparison of the two results is complicated because their sample is dominated by the bar population (with G1 characteristics) in low-latitude fields.
Since the current view on the 3D mapping of the Milky Way bulge is based on the previous interpretation of the double RC phenomenon \citep{ref12}, our new result calls for a major revision in our quest for the structure of the Milky Way bulge. 
Our result would suggest that the Milky Way has a composite bulge with a bar embedded in a bulge component with GC origin. 
In this paradigm, the variation of the RC luminosity function with Galactic longitude and latitude is well explained in the multiple population scenario, which is also not inconsistent with the observed kinematics \citep{ref3,ref5}. 
Note that the stellar density in the claimed faint X-shaped structure from WISE residual map \citep{ref24} is way too low to be observed as the double RC \citep{ref26,ref25}.
\par
Because the classical bulge component of the Milky Way bulge has long been considered as the nearest early-type galaxy (ETG) for which we can resolve into stars, our discovery in the Milky Way bulge would further imply that ETGs would be similarly consist of stars having GC origins. 
In this respect, it is interesting to find that most massive ETGs and their GCs are Na and CN enhanced \citep{ref27,ref28}. 
Further investigations with Gaia trigonometric parallax distances and high-resolution spectroscopy for the bulge stars in the double RC would undoubtedly help to strengthen our new results and the implications drawn from them.

\acknowledgments
We are grateful to John Norris for many helpful discussions and suggestions. We also thank the anonymous referee for constructive comments, and the staff of LCO for their support during the observations. This work was supported by the National Research Foundation of Korea (grants 2017R1A6A3A11031025, 2017R1A2B3002919, and 2017R1A5A1070354).


\end{document}